\documentclass[conference]{IEEEtran}
\IEEEoverridecommandlockouts

\usepackage{amssymb}
\usepackage[cmex10]{amsmath}
\usepackage{stfloats}
\usepackage{graphicx}
\usepackage{subfigure}
\usepackage{tabularx}
\usepackage{epsfig,epsf,color,balance,cite}
\usepackage{verbatim}
\usepackage{url}
\usepackage{bm}
\usepackage{booktabs}
\usepackage[top=0.75in, bottom=1.14in, left=0.64in, right=0.64in]{geometry}

\usepackage{algorithm}
\usepackage{algorithmic}

\usepackage{color}
\definecolor{myc1}{rgb}{0,0,0}

\begin{document}

\title{ 
 Resource Allocation for
 Green Probabilistic Semantic Communication with Rate Splitting
}

\author{
\IEEEauthorblockN{Ruopeng Xu\IEEEauthorrefmark{1},
                  Zhaohui Yang\IEEEauthorrefmark{1}\IEEEauthorrefmark{2},
                  Zhouxiang Zhao\IEEEauthorrefmark{1},
                  Qianqian Yang\IEEEauthorrefmark{1}\IEEEauthorrefmark{2},
                  and Zhaoyang Zhang\IEEEauthorrefmark{1}\IEEEauthorrefmark{2}
                 }
	\IEEEauthorblockA{
			$\IEEEauthorrefmark{1}$College of Information Science and Electronic Engineering, Zhejiang University, Hangzhou, China\\
			$\IEEEauthorrefmark{2}$Zhejiang Provincial Key Laboratory of Info. Proc., Commun. \& Netw. (IPCAN), Hangzhou, China\\
			E-mails:
   \{ruopengxu, yang\_zhaohui, zhouxiangzhao, qianqianyang20, ning\_ming\}@zju.edu.cn
		}
\thanks{This work was supported by the National Key R\&D Program of China (Grant No. 2023YFB2904804), National Natural Science Foundation of China (NSFC) under Grants 62394292, 62394290.}
\vspace{-2em}
}

\maketitle

\begin{abstract}
In this paper, the energy efficient design for probabilistic semantic communication (PSC) system with rate splitting multiple access (RSMA) is investigated. Basic principles are first reviewed to show how the PSC system works to extract, compress and transmit the semantic information in a task-oriented transmission. Subsequently, the process of how multiuser semantic information can be represented, compressed and transmitted with RSMA is presented, during which the semantic compression ratio (SCR) is introduced to directly measure the computation overhead in a transmission task, and communication overhead is indirectly described as well. Hence, the problem of wireless resource allocation jointly considering the computation and communication consumption for the PSC system with RSMA is investigated. Both conventional wireless resource constraints and unique constraints on semantic communication are considered to maximize the energy efficiency (EE). Simulation results verify the effectiveness of the proposed scheme.
\end{abstract}

\IEEEpeerreviewmaketitle

\section{Introduction} \label{Introduction}
Colossal data generated by the rapid-developing emerging applications\cite{10024766}, such as autonomous vehicles, satellite communication and metaverse are in the need of being transmitted in a high-rate and low-latency way. Conventional communication system model proposed by Shannon in 1948 focusing on maximizing the transmitting rate is getting more and more close to its theoretical limit. In addition to this, the increasingly precious spectrum resources and hardware devices are also forcing us to find a new communication paradigm that can solve the problem. 

\begin{figure*}[t]
\centering
\includegraphics[width=0.95\linewidth]{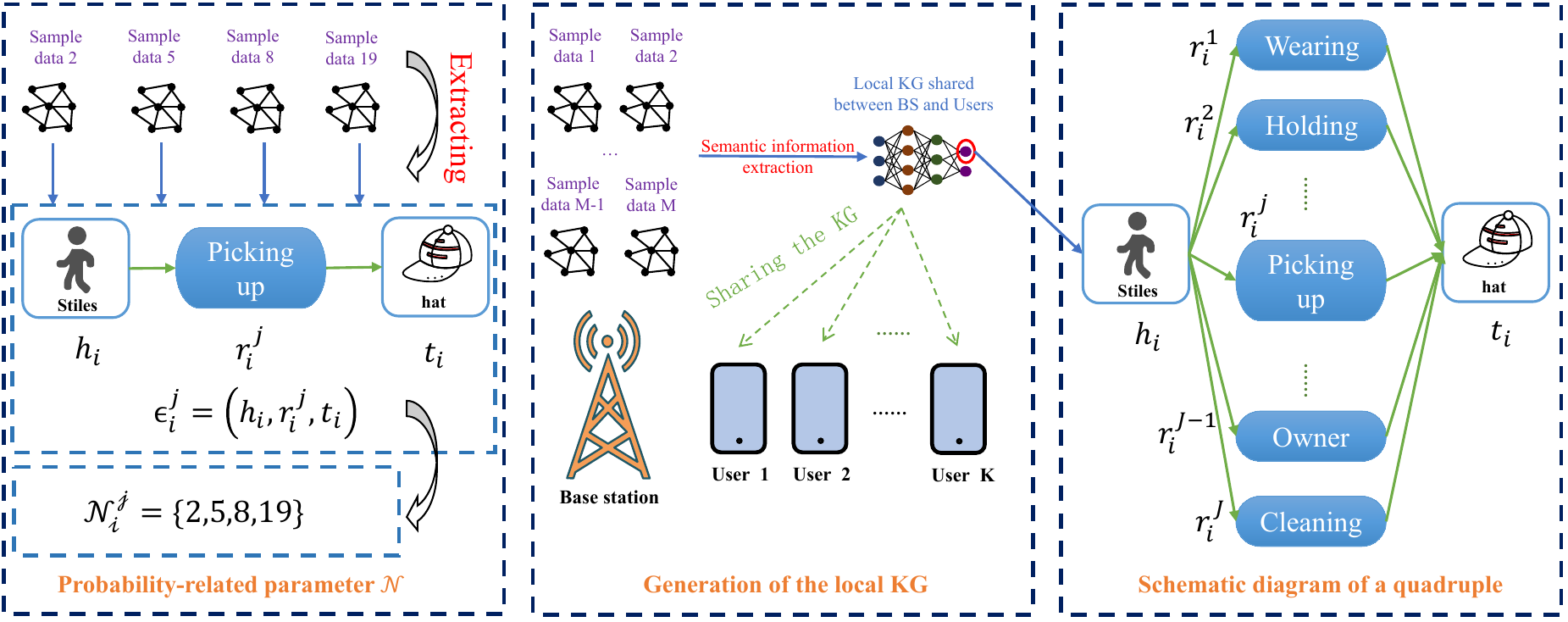}
\caption{A schematic diagram of PSC model} 
\label{fig1}
\end{figure*}

Semantic communication is recognized as a promising paradigm for the next generation of wireless communication \cite{yang2023energy}, which pays attention to figuring out the core point of one communication task first, and then execute the transmission. 
Semantic communication is not a brand new concept, and the significance of it in communication has already been emphasized by Shannon and Weaver. Although there is no unified definition of semantic communication and its related concepts, in recent years, with the continuous development of artificial intelligence (AI) technology and computing power, an abundance of researches have been conducted on various of semantic communication system. Due to the lack of a general system model for semantic communication, authors in \cite{zhang2022toward} proposed semantic base (Seb) as a new semantic representation framework pointing out that semantic information can be characterized by feature parameters in high-dimensional space. Considering the trend that future communication will not take place only between humans and humans, but also happen between humans and machines and even machines and machines, authors in \cite{niu2022paradigm} provided a block diagram of a semantic communication system for both human-type communications and machine-type communications.

It has been widely acknowledged that knowledge graph (KG) can effectively represent complex information with the explosive growth of AI and big data by accumulating and conveying knowledge of the real world. One of the typical techniques represent a KG as a set of triples, where each triple  $(h, r, t)$ consists of two entities $h$ and $t$ connected by a relation $r$\cite{rosso2020beyond}. How to create a KG with the local knowledge base (KB) is one of the valuable issues on AI area, and it provides a potential approach to semantic information extraction\cite{10233741}. In \cite{wang2022performance}, authors leveraged the shared KB and KG to enhance the semantic communication system to extract and exploit the meaning behind digital bits, and jointly optimized semantic accuracy and semantic completeness to measure the quality of semantic communication. In \cite{zhou2022cognitive}, authors proposed a cognitive semantic communication framework and a simple solution to semantic information detection by exploiting the KG and its triples as semantic symbols, respectively.

Although \cite{zhang2022toward,niu2022paradigm,wang2022performance,zhou2022cognitive} take the shared KB and KG into consideration, utilizing them as useful tools to extract and represent the semantic information, none of them consider the computing energy consumption and its connection with the communication consumption of a semantic communication network. Whether it is the requirement of low-carbon technique or the constraints of the available power for wireless communication resources, it is necessary to fully consider the power consumption in various aspects when designing a semantic communication system, such as the compression calculation of semantic information and the transmission of semantic information. Since information and communication technology accounts for $5\%$ consumption all over the world\cite{zhang2016energy}, great attention is supposed to be paid to maximizing the energy efficiency (EE) as much as possible, enabling the semantic communication system to achieve a green status in terms of energy usage.

Rate splitting multiple access (RSMA), introduced in \cite{rimoldi1996rate,clerckx2016rate,liu2020rate,clerckx2024multiple,9145189}, divides the messages intended to be transmitted into shared part and private part. Users at the receiver utilize successive interference cancellation (SIC) technology to acquire partly decoded interference \cite{9382277,10038476,9451194,9831440}. Inspired by the likelihood of overlapping transmission tasks among multiple users in task-oriented semantic communication and the fact\cite{zhou2021rate} that rate splitting multiple access can improve energy efficiency, we aim to use rate splitting multiple access to assist in building a multiuser green semantic communication system.

Motivated by the aforementioned observations, in this paper, we proposed a downlink RSMA semantic communication system with multiple users utilizing the probabilistic semantic communication (PSC) model as we forehead proposed in \cite{zhao2023semantic} to extract and represent the semantic information and integrate the computation and communication processes to formulate an optimization problem that maximizes EE in order to meet the green requirement. The key contributions are listed as follows: 
\begin{itemize}
\item We further extend the PSC model from the single BS and single user to the multiuser scenario. Simultaneously considering the overlapping nature of tasks needed to be transmitted for multi-user semantic communication, we introduce RSMA to enhance the EE of the system.
\item We introduce the shared information semantic compression ratio and private information semantic compression ratio to jointly consider the computational overhead in semantic compression and the communication overhead in transmission, which makes the concept of green communication more comprehensive.
\item In the optimization goal of maximizing EE, the computational power is introduced into the total power consumption of the system. Regarding the constraints of the optimization problem, a comprehensive consideration is given to the limited computational capabilities and transmission power of BS caused by limited wireless communication resources, user latency requirements, and unique constraint of semantic communication. 
\end{itemize}


The rest of this paper is organized as follows. The system model and problem formulation are described in Section \ref{System Model and Problem Formulation}. Simulation results and according analyses are presented in Section \ref{Simulation Results and Analysis}. Conclusions are drawn in Section \ref{Conclusion}.

\section{System Model and Problem Formulation} \label{System Model and Problem Formulation}
In our previous work, we have established a PSC model with single BS and single user situation. For both reading coherence and the convenience of the subsequent development of the article, we reiterate some basic but important definitions above all, and more specific details about the PSC model and the probability enhanced information compression method can be found in \cite{zhao2023semantic}. 

\subsection{Review of PSC model}
The basic elements of the local KG, which is also as the semantic information of the semantic communication, generated by sample data $\mathcal{D}$ are the triples in the form of:
\begin{equation}
    \begin{aligned} 
    \epsilon_i^j = (h_i,r_i^j,t_i),
    \end{aligned}
\label{Triple}
\end{equation}
where $\epsilon_i^j$ stands for the triple consisting of the entity pair $h_i$ and $t_i$ with the relation $r_i^j$ pointing from $h_i$ to $t_i$, as shown in the left column in Fig.~\ref{fig1}. For one specific entity pair $h_p$ and $t_p$, there are probably more than one relation, as shown in the right column in Fig.~\ref{fig1}. If introduce the probability-related parameter $\mathcal{N}$ into the triples sharing the same entity pair, one quadruple set can be established in the form of:
\begin{equation}
    \begin{aligned} 
    \delta_p = \left\{h_p,\left[\left(r_p^1, \mathcal{N}_p^1 ),\dots,(r_p^q,\mathcal{N}_p^q),\dots,(r_p^Q,\mathcal{N}_p^Q\right)\right],t_p\right\},
    \end{aligned}
\label{Quadruple}
\end{equation}
where $r_p^q$ stands for the $q$-th relation in between $h_p$ and $t_p$, and set $\mathcal{N}_p^q$ is a number set consisting of the serial number of data samples that can extract the triple $\epsilon_p^q$. Briefly speaking, a PSC model can be described as shown in Fig.~\ref{fig1}.

In addition, the probability of a triple $\epsilon_i^j$ in the KG can be defined as:
\begin{equation}
    \begin{aligned} 
    p(\epsilon_i^j) = \frac{card\left(\mathcal{N}_i^j\right)}{\sum_{j=1}^J card\left(\mathcal{N}_i^j\right)} = p_i^j,
    \end{aligned}
\label{ProbabilityDefination}
\end{equation}
and the conditional probability of $\epsilon_i^j$ when known $\epsilon_a^b$ can be defined as:
\begin{equation}
    \begin{aligned} 
    p(\epsilon_i^j|\epsilon_a^b) = \frac{card\left(\mathcal{N}_i^j \cap \mathcal{N}_a^b \right)}{card\left[\left(\cup_{b=1}^B \mathcal{N}_a^b\right) \cap \mathcal{N}_i^j\right]} = p_{i_a^b}^j,
    \end{aligned}
\label{ConditionalProbabilityDefination}
\end{equation}

where $B$ is the total number of the relations pointing from $h_a$ to $t_a$ in the local KG. 
In following of this section, we first tell the story about how to represent the semantic information of multiple users based on PSC, and in addition build a downlink RSMA semantic communication system whose computation overhead and communication overhead are jointly considered by the parameter we proposed, semantic compression ratio (SCR), of the PSC model. Last but not least, we formulate the optimization problem of maximizing the EE of the system under some specific constraints.

\subsection{Multiuser Semantic Information Representation}
\begin{figure}[t]
\centering
\includegraphics[width=1\linewidth]{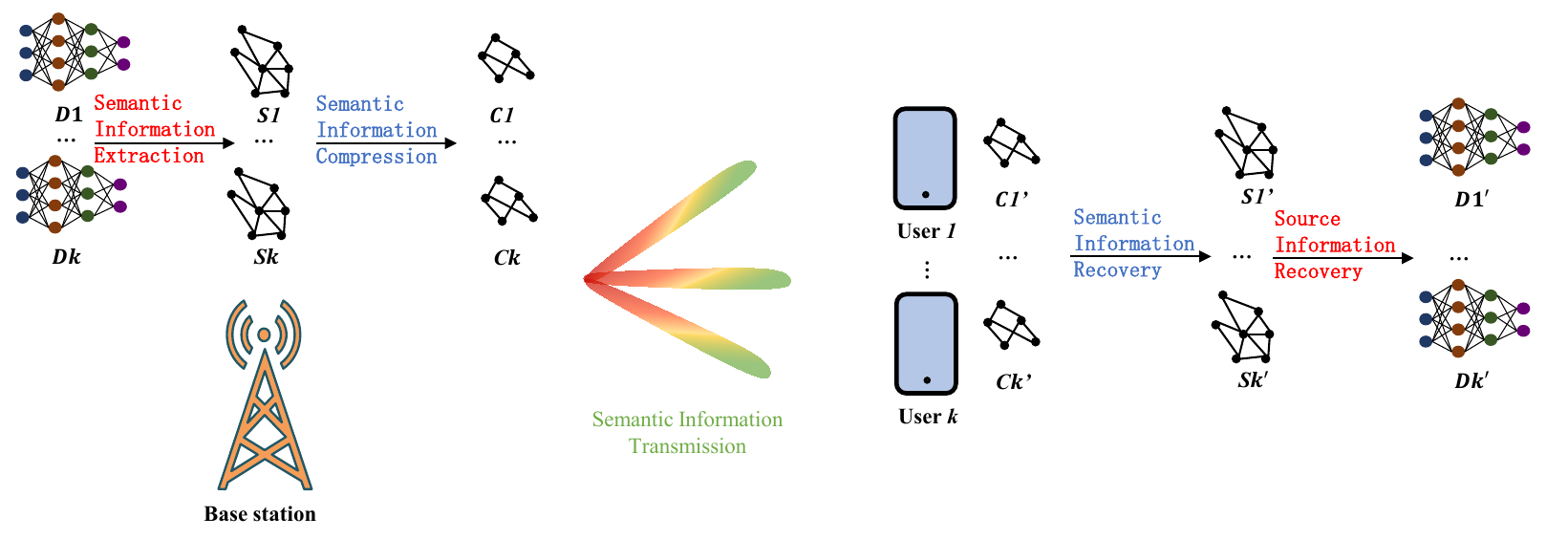}
\caption{Multiuser semantic communication system based on PSC} 
\label{fig2}
\end{figure}

\addtolength{\topmargin}{0.02in}
Consider the situation, shown in Fig.~\ref{fig2}, that the BS has already generated a local KG before. Assume that there are one single BS with multiple antennas and $K$ users with single antenna, respectively, and every user has a great amount of data to transmit. After being extracted by the BS, the data to be transmitted from user $1$ to user $K$ in the form of sets of triples can be represented as:
\begin{equation}
\left\{
    \begin{aligned} 
        & \mathcal{S}_1 = \{\epsilon_{11}, \epsilon_{12}, \dots, \epsilon_{1k_1}\},\\
        & \mathcal{S}_2 = \{\epsilon_{21}, \epsilon_{22}, \dots, \epsilon_{2k_2}\},\\
        & \vdots\\
        & \mathcal{S}_K = \{\epsilon_{K1}, \epsilon_{K2}, \dots, \epsilon_{Kk_K}\},
    \end{aligned}
\right.
\label{MultiuserSemanticInformation}
\end{equation}
\addtolength{\topmargin}{0.04in}
where $\mathcal{S}_i$ is the semantic information set of user $i$ and $\epsilon_{ij}$ means that it is the $j$-th triple of user $i$. Then, compare all triples of $K$ users, find the triples shared by every user, and merge them into common triples, denoted as shared triple $\varepsilon$, i.e. shared information, and the rest triples stay still and are denoted as private triples, i.e. private information. Last but not least, all the semantic triples are sorted out as:
\begin{equation}
    \begin{aligned} 
        & S = \{\varepsilon_1, \varepsilon_2, \dots, \varepsilon_M, \epsilon_{11}, \dots, \epsilon_{1k'_1}, \epsilon_{21},\dots, \epsilon_{K1}, \epsilon_{Kk'_K}\},\\
    \end{aligned}
\label{SharedTripleSet}
\end{equation}
where $M$ denotes that there are $N$ shared triples of all the $K$ users, and $k'_i$ means that there are $k'_i$ private triples of user~$i$.

\subsection{Semantic Information Compression}
Assume that there are $\Phi$ quadruple sets can be constructed from the local KG, and for every quadruple set $\delta_i$, there are $j_i$ relations pointing from $h_i$ to $t_i$. Utilizing the definitions of the probability, we first calculate the probability of all the triples and form the probability matrix with no prior knowledge. In order to ensure that each row of the probability matrix can represent a quadruple set, the rows corresponding to the quadruples with less relations are padded with zeros. Hence the probability matrix with no prior knowledge can be denoted as:
\begin{equation}
    \begin{aligned}
      \mathbf{P} =
      \begin{bmatrix}
      p_1^1 & p_1^2 & \dots & p_1^{j_1} & \dots & 0\\
      \vdots & & & & & \vdots\\
      p_m^1 & p_m^2 & \dots & \dots & \dots & p_m^{j_m}\\
      \vdots & & & & & \vdots\\
      p_\Phi^1 & p_\Phi^2 & \dots & p_\Phi^{j_\Phi} & \dots & 0\\
      \end{bmatrix}
      \in \mathcal{R}^{\Phi \times j_m}
    \end{aligned}
\label{ProbabilityMatrixwithnoKnowledge}
\end{equation}
where $j_m = {\max\{j_i\}}_{i=1}^{\Phi}$. 

For the semantic information set $\mathcal{S}$ in a specific transmission task, traverse every triple of the set. If the probability corresponding to a triple $\epsilon_i^j$/$\varepsilon_i$ is the value of the matrix $\mathbf{P}$ and this value is the maximum value of a certain row, omit the relation of the triple during the subsequent transmission, and record it as a degenerate triple $o_i^j$/$o_i$.

After the first round of traversal, some triples in the set $\mathcal{S}$ have completed degradation. Taking the initial triple corresponding to the degenerate triple in $\mathcal{S}$ as a condition, a probability matrix similar to the matrix $\mathbf{P}$ can be calculated. For example, if there are three triples $\{\epsilon_1^1, \epsilon_3^4, \varepsilon_2\}$ are judged to be degenerated in the first compression round, and they are denoted as $\{o_1^1, o_3^4, o_2\}$. Taking $o_1^1$ as the condition, recalculate the probability of each triple in each quadruple set to obtain the one-dimensional conditional probability matrix $\mathbf{P}_1^{o_1^1}$, which can be expressed as:
\begin{equation}
    \begin{aligned}
      \mathbf{P}_1^{o_1^1} =
      \begin{bmatrix}
      p_{1_1^1}^1 & p_{1_1^1}^2 & \dots & p_{1_1^1}^{j_1} & \dots & 0\\
      \vdots & & & & & \vdots\\
      p_{m_1^1}^1 & p_{m_1^1}^2 & \dots & \dots & \dots & p_{m_1^1}^{j_m}\\
      \vdots & & & & & \vdots\\
      p_{\Phi_1^1}^1 & p_{\Phi_1^1}^2 & \dots & p_{\Phi_1^1}^{j_\Phi} & \dots & 0\\
      \end{bmatrix}
      \in \mathcal{R}^{\Phi \times j_m}
    \end{aligned}
\label{OneDimentionalProbabilityMatrixwithnoKnowledge}
\end{equation}
where $1$ and $o_1^1$ of $\mathbf{P}_1^{o_1^1}$ mean that $\mathbf{P}_i^{o_1^1}$ is a one-dimensional matrix and it takes $o_1^1$ as the prior knowledge, respectively. Then traverse the triples in the set that are not degenerate. If the conditional probability value corresponding to a triple is in the matrix $\mathbf{P}_1^{o_1^1}$ and is the maximum value of a row of the matrix, omit the relation and record it as degenerate as well. Generate new one-dimensional conditional probability matrices $\mathbf{P}_1^{o_3^4}$ and $\mathbf{P}_1^{o_2}$ based on ${o_3^4}$ and $o_2$, respectively, re-traverse the non-degraded triples in the set according to the same omission rules, and the second compression round is completed.

Subsequently, by constructing two-dimensional probability matrices, three-dimensional probability matrices and even N-dimensional probability matrices, the third, the fourth, and even the $(N+1)$ round of semantic information compression can be completed. If there are $Q_1$ triples omitted in the first round, there will be $C_{Q_1}^1$ one-dimensional matrices to be calculated in the second round, and without loss of generality, if there are $Q_{N}$ triples omitted in the $N$-th round, there will be $C_{\sum_{i=1}^N Q_N}^{N}$ $N$-dimensional matrices to be calculated in the $(N+1)$-th round.

It is easy to know that omitting the transmission of the relation in triples can reduce communication overhead, but it will introduce computational overhead in some degree. According to the compression rules introduced above, it can be found that as the number of compression rounds increases, the probability matrices that need  to be calculated in each round will increase nonlinearly, and the triples that can be degraded in each round will not increase proportionally, and may even will decrease round by round.

For the convenience of description, we introduce SCR defined as Eq.~\eqref{SCR}
\begin{equation}
    \begin{aligned} 
         \Omega &= \dfrac{\sum_{i=1}^N Q_i}{M+\sum_{i=1}^K k'_i},
    \end{aligned}
\label{SCR}
\end{equation}
where $Q_i$ denotes the number of the degenerate triples in the $i$-th round of compression, and $N$ means it is the SCR after N-round compression, the denominator stands for the total number of the triples need to transmit as set in (\ref{SharedTripleSet}). Then, we describe the relationship between the cost of introduced computational and reduced communication and the SCR.

\begin{figure}[t]
\centering
\includegraphics[width=1\linewidth]{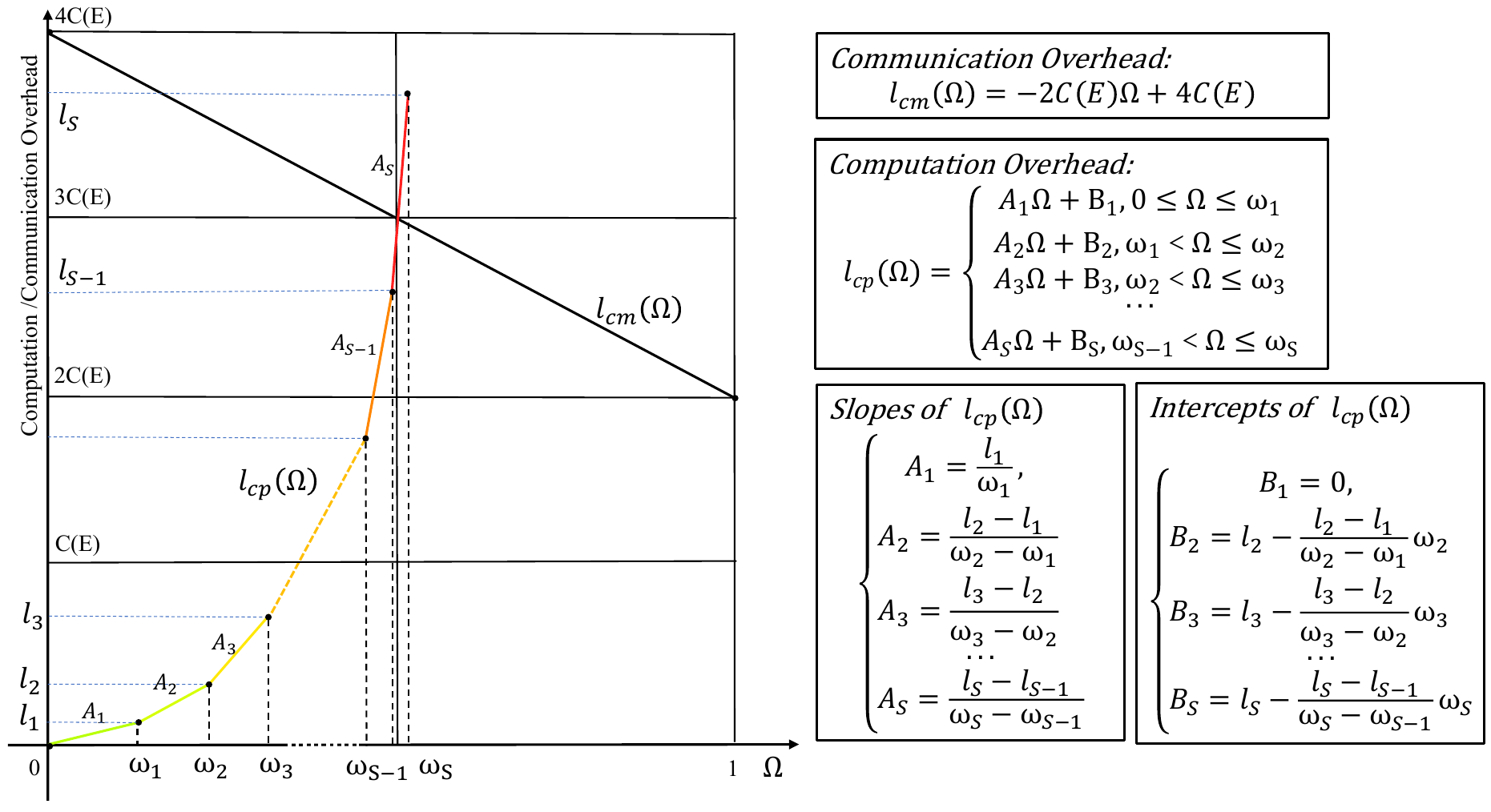}
\caption{Mapping relations between $\Omega$ and $l_{cp}(\Omega)$/$l_{cm}(\Omega)$} 
\label{fig3}
\end{figure}

Equal-length coding is used to encode triples. Considering that the relation $r_i^j$ of a triple $\epsilon_i^j$ contains a larger amount of information, for $\epsilon_i^j$, half of the code word is used to encode $r_i^j$, and the remaining half of the code word is used to encode $h_i$ and $t_i$. For convenience of description, denote that the code word length of $h_i$ in a non-degraded triple is $E$, the code word length of $\epsilon_i^j$ is $2E$, and the code word length of $t_i$ is also $E$. $C(\cdot)$ reflects the length of code word to communication overhead $l_{cm}(\Omega)$. Then the relationship between the $\Omega$ and computation overhead $l_{cp}(\Omega)$ /communication overhead $l_{cm}(\Omega)$ can be given in Fig.~\ref{fig3}.

\subsection{Downlink RSMA Semantic Communication Model} 
Consider the feature that the triples are divided into shared ones and private ones, we utilize RSMA for the semantic information transmission of the multiple users. 

For the signal $\mathbf{x}$ to be transmitted in the BS can be written as:
\begin{equation}
    \begin{aligned} 
         \mathbf{x} = \sqrt{p_0}\mathbf{w_0}m_0 + \sum_{i=1}^{K} \sqrt{p_i}{\mathbf{w}}_im_i ,\\
    \end{aligned}
\label{x}
\end{equation}
where ${\mathbf{w}}_0$ is the transmit beamforming of the shared information $m_0$ consists of the shared triples, ${\mathbf{w}}_i$ is the transmit beamforming of the private information $m_i$ consists of the private triples of user $i$, $p_0$ stands for the transmit power allocated to shared message, and $p_i$ stands for the transmit power allocated to private message of user $i$.

For user $i$, the received message can be written as:
\begin{equation}
    \begin{aligned} 
         y = \mathbf{h}_i^H\mathbf{x} + n_i ,\\
    \end{aligned}
\label{y}
\end{equation}
where $\mathbf{h}_i^H$ stands for the channel between the BS and user $i$, and $n_i$ the Gaussian noise with power $\sigma^2$. Then the rate of user $i$ to decode the shared message can be given as:
\begin{equation}
    \begin{aligned} 
         s_i = B{\log}_2\left(1+\dfrac{p_0 |\mathbf{h}_i^H \mathbf{w}_0|^2}{\sum_{j=1}^K p_j |\mathbf{h}_i^H \mathbf{w}_j|^2+\sigma^2}\right) ,\\
    \end{aligned}
\label{SharedMessageRate}
\end{equation}
where $B$ is the bandwidth of BS. Note that in addition to shared messages and private message of user $i$, other private messages are interference to user $i$, the rate of user $i$ to decode the private message can be presented as:
\begin{equation}
    \begin{aligned} 
         q_i = B{\log}_2\left(1+\dfrac{p_i |\mathbf{h}_i^H \mathbf{w}_i|^2}{\sum_{j=1,j\neq i}^K p_j |\mathbf{h}_i^H \mathbf{w}_j|^2+\sigma^2}\right) .\\
    \end{aligned}
\label{PrivateMessageRate}
\end{equation}

To ensure all the users can decode the shared message successfully, the rate of the shared message is supposed to be set as\cite{mao2018rate}:
\begin{equation}
    \begin{aligned} 
         s_0 = \mathop{\min}_{i\in K}s_i .\\
    \end{aligned}
\label{so}
\end{equation}

\subsection{Problem Formulation}
For a transmission task involving $K$ users and one BS, assume that there are $M$ shared triples and $k'_i$ private triples of user $i$, i.e. there are $\sum_{i=i}^K M_i$ private triples in total, in the semantic set $\mathcal{S}$ as Eq.~\eqref{MultiuserSemanticInformation} shows. After certain rounds of compression, assume the SCR of shared triples and private triples of user $i$ are $\Omega_0$ and $\Omega_i$, respectively. Then we can obtain the compression vector $\bm{\Omega} = [\Omega_0, \Omega_1, \dots,\Omega_K]^T$.

For user $i$, jointly consider the compression for the semantic information in the BS and transmission from the BS to user $i$, and the time delay can be presented as:
\begin{equation}
    \begin{aligned} 
         \tau_i = \dfrac{l_{cp}(\Omega_{i})}{f_i} + \mathop{\max}\left\{\dfrac{l_{cm}(\Omega_0)}{s_0},\dfrac{l_{cm}(\Omega_{pi})}{q_i}\right\} ,\forall i \in \mathcal{K}\\
    \end{aligned}
\label{Timedelay}
\end{equation}
where $l_{cp}(\cdot)$ and $l_{cm}(\cdot)$, as shown in Fig.~\ref{fig3}, are the mapping relation from SCR to computation overhead and communication overhead, respectively, $f_i$ is the computing capability allocated for user $i$, and $f_0$ is the computing capability allocated for compressing shared message. 
And for shared message, its time delay can be described as:
\begin{equation}
    \begin{aligned} 
         \tau_0 = \dfrac{l_{cp}(\Omega_{0})}{f_0} + \dfrac{l_{cm}(\Omega_{0})}{s_0}.\\
    \end{aligned}
\label{Timedelay2}
\end{equation}

Assume every code word contain $R$ bit information, the $l_{cm}(\Omega)$ can be further expressed as:
\begin{equation}
    \begin{aligned} 
         l_{cm}(\Omega) = 2C(E)(2-\Omega) = 2RE(2-\Omega) ,\Omega\in[0,1].\\
    \end{aligned}
\label{communicationOverhead}
\end{equation}

In order to ensure that semantic information has a satisfying accuracy between the transmitter and the receiver, the semantic accuracy parameter $A(S,S')$ is introduced:
\begin{equation}
    \begin{aligned} 
         A_i(\mathcal{S},\mathcal{S}'_i) = \dfrac{\sum_{i=1}^I {\min}\left\{\sigma(\mathcal{S},\epsilon_i),\sigma(\mathcal{S}'_i,\epsilon_i)\right\}}{\sum_{i=1}^I {\min}\left\{\sigma(\mathcal{S},\epsilon_i)\right\}},\\
    \end{aligned}
\label{SemanticAccuracy}
\end{equation}
where $I$ means the total number of the different triples in the semantic information set $\mathcal{S}$ is $I$, $\mathcal{S}'_i$ is the semantic information received and recovered by user $i$, and $\sigma(\mathcal{S},\epsilon_i)$ is the number of occurrences of $\epsilon_i$ in $\mathcal{S}$.

When SCR is determined, the local computation energy consumed to compress messages can be given by:
\begin{equation}
    \begin{aligned} 
         E_{cpi} = \xi l_{cp}(\Omega_i){f_i}^2, i = 0,1\dots,K\\
    \end{aligned}
\label{ComputationEnergyforShared}
\end{equation}
where $\Omega_0$ stands for the computation energy for compressing shared message, $\Omega_i$ stands for the computation energy for compressing private message of user $i$, and $\xi$ is a constant coefficient to measure the effective switched capacitance.

The communication energy caused by sending shared information can be presented as:
\begin{equation}
    \begin{aligned} 
         E_{cm0} = \dfrac{l_{cm}(\Omega_0)}{s_0}p_0.\\
    \end{aligned}
\label{CommunicationnEnergyforShared}
\end{equation}

The communication energy caused by sending private message of user $i$ can be presented as:
\begin{equation}
    \begin{aligned} 
         E_{cmi} = \dfrac{l_{cm}(\Omega_{i})}{q_i}p_i.\\
    \end{aligned}
\label{CommunicationnEnergyforPrivate}
\end{equation}

Subsequently, if define the energy efficiency as the ratio of the number of bits in one transmission task to all energy consumed in completing a communication task, then, when given the semantic set $\mathcal{S}$ it can be given as:
\begin{equation}
    \begin{aligned} 
         EE = \dfrac{4RE(M+\sum_{i=1}^K k'_i)}{\sum_{i=0}^K (E_{cpi}+E_{cmi}) + E_0},\\
    \end{aligned}
\label{EE}
\end{equation}
where $E_0$ is a constant standing for the energy consumed by the circuit.

Based on what we have talked all, we can construct the following joint optimization problem
\begin{subequations}\label{OptimizationProblem}
    \begin{align} 
        \mathop{\max}_{\mathbf{p},\mathbf{f},\Omega_s,\Omega_p} \quad & EE ,\tag{\ref{OptimizationProblem}}\\
         \textrm{s.t.} \qquad & A_i(\mathcal{S},\mathcal{S'_i}) \geq A_{min}, \forall i \in \mathcal{K}, \\
         & \sum_{i=0}^K p_i \leq P_{max},\\
         & \sum_{i=0}^K f_i \leq F_{max},\\
         & \tau_i \leq T_{max}, \forall i \in \mathcal{K} \cup \{0\},\\
         & f_i,p_i \geq 0, \forall i \in \mathcal{K} \cup \{0\},\\
         & 0 \leq \Omega_i \leq 1,\forall i \in \mathcal{K} \cup \{0\}\\
         & \|\mathbf{w}_i\| = 1, \forall i \in \mathcal{K} \cup \{0\},
    \end{align}
\end{subequations}
where $\mathbf{f}=[f_0,f_1,\dots,f_K]^T$, $\mathbf{w} = [w_0,w_1,\dots,w_K]^T$, $\mathbf{p}=[p_0, p_1, \dots, p_K]^T$, $\mathcal{K} = \{1,2,\dots,K\}$, $A_{min}$ is the minimum value of the semantic accuracy, $P_{max}$ is the maximum transmission power in the BS, $F_{max}$ is the maximum computation capacity in the BS, and $T_{max}$ is the maximum time delay acceptable to all the users. The objective of the
optimization problem is to maximize the energy efficiency  of the semantic communication system while obeying all the constraints. 

To figure out the solution to problem ~\eqref{OptimizationProblem}, SCRs of shared triples and private triples are supposed to be determined by the BS utilizing the training samples to find the least overhead jointly considering computation and communication when given a specific semantic transmission task, and gradient descent optimization method is leveraged to optimize $\mathbf{p}$ and $\mathbf{f}$ with given $\bm{\Omega}$.

\section{Simulation Results and Analysis} \label{Simulation Results and Analysis}
During the simulation process, keep most of the parameter settings the same as in \cite{zhao2023semantic} , where the minimum semantic accuracy $A_{min} = 0.9$, the maximum acceptable time delay $T_{max} = 1$ $\mathrm{ms}$, $P_{max} = 30$ $\mathrm{dBm}$, and $F_{max} = 3 \times 10^9$.

To evaluate the performance of the proposed PSC system with RSMA under the multiuser scenario, we develop the comparison among the system optimized with the proposed optimization method, labeled as `Proposed', with no compression process, labeled as `$\Omega_i = 0$' and with SCR not in the ideal situation, labeled as `$\Omega_i = 0.3$', i.e., all the compression ratios in the compression vector $\bm{\Omega}$ are set to be equal to $0.3$. In the simulation of the 'Proposed', the compression vector $\bm{\Omega}$ is initially determined by the BS utilizing the training samples to find the empirical least overhead jointly considering computation and communication, and optimized in some range to find the practical least overhead. The baseline with no compression, `$\Omega_i = 0$', optimizes the allocation of communication and computation resources only. The baseline, `$\Omega_i = 0.3$', optimizes the allocation of communication and computation resources with compression in imperfect compression.

\begin{figure}[t]
\centering
\includegraphics[width=1\linewidth]{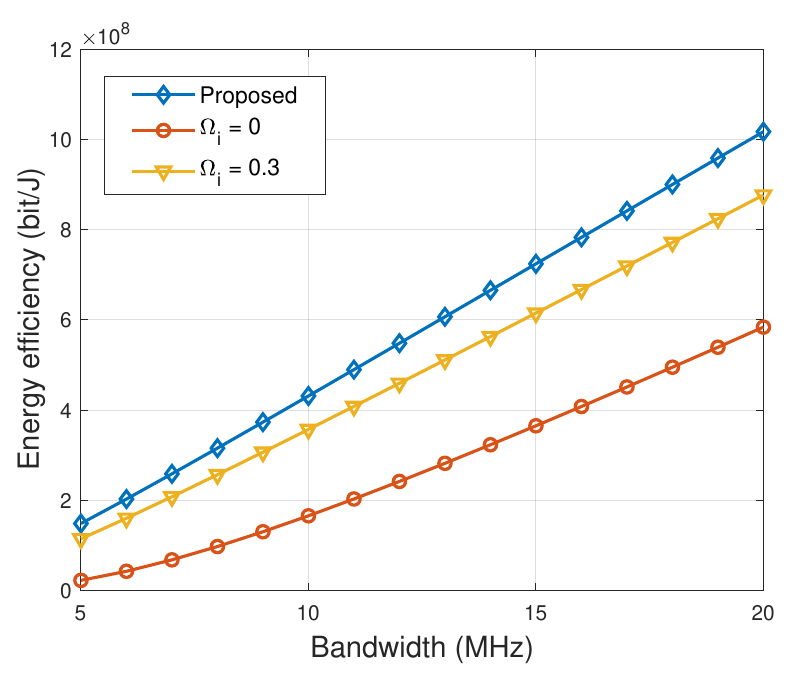}
\caption{Energy efficiency vs. bandwidth of the BS} 
\label{fig4}
\end{figure}

Fig.~\ref{fig4} depicts the changing trend of the energy efficiency of the PSC system with respect to the bandwidth of the BS. As the bandwidth of the local BS increases, the energy efficiency of all compression degree in the system also increases due to the lower power requirement of the communication. Nevertheless, the `Proposed' algorithm demonstrates a larger increase trend compared with the other two baselines, which illustrates a better performance of the proposed system. Furthermore, the advantage shown as the bandwidth of the BS increased is that the EE of `Proposed' is constantly higher than the other two baselines. The outcome highlights that not only the system performs better under a better condition, but also remains better under a poorer one. 

\section{Conclusion} \label{Conclusion}
In this paper, we have developed the PSC system from the single user scenario to multiuser scenario with RSMA. After sequentially modeling the representation, compression and transmission of multiuser semantic information, we constructed an optimization problem to maximize the energy efficiency of the system on the basis of satisfying certain optimization constraints. Simulation result shows that the proposed system exhibits better performance than no semantic compression and inappropriate semantic compression ratio selection.

\bibliographystyle{IEEEtran}
\bibliography{main}

\begin{thebibliography}{10}
\providecommand{\url}[1]{#1}
\csname url@samestyle\endcsname
\providecommand{\newblock}{\relax}
\providecommand{\bibinfo}[2]{#2}
\providecommand{\BIBentrySTDinterwordspacing}{\spaceskip=0pt\relax}
\providecommand{\BIBentryALTinterwordstretchfactor}{4}
\providecommand{\BIBentryALTinterwordspacing}{\spaceskip=\fontdimen2\font plus
\BIBentryALTinterwordstretchfactor\fontdimen3\font minus \fontdimen4\font\relax}
\providecommand{\BIBforeignlanguage}[2]{{%
\expandafter\ifx\csname l@#1\endcsname\relax
\typeout{** WARNING: IEEEtran.bst: No hyphenation pattern has been}%
\typeout{** loaded for the language `#1'. Using the pattern for}%
\typeout{** the default language instead.}%
\else
\language=\csname l@#1\endcsname
\fi
#2}}
\providecommand{\BIBdecl}{\relax}
\BIBdecl

\bibitem{10024766}
W.~Xu, Z.~Yang, D.~W.~K. Ng, M.~Levorato, Y.~C. Eldar, and M.~Debbah, ``Edge learning for {B5G} networks with distributed signal processing: Semantic communication, edge computing, and wireless sensing,'' \emph{IEEE J. Sel. Topics Signal Process.}, vol.~17, no.~1, pp. 9--39, Jan. 2023.

\bibitem{yang2023energy}
Z.~Yang, M.~Chen, Z.~Zhang, and C.~Huang, ``Energy efficient semantic communication over wireless networks with rate splitting,'' \emph{IEEE Journal on Selected Areas in Communications}, vol.~41, no.~5, pp. 1484--1495, 2023.

\bibitem{zhang2022toward}
P.~Zhang, W.~Xu, H.~Gao, K.~Niu, X.~Xu, X.~Qin, C.~Yuan, Z.~Qin, H.~Zhao, J.~Wei \emph{et~al.}, ``Toward wisdom-evolutionary and primitive-concise 6g: A new paradigm of semantic communication networks,'' \emph{Engineering}, vol.~8, pp. 60--73, 2022.

\bibitem{niu2022paradigm}
K.~Niu, J.~Dai, S.~Yao, S.~Wang, Z.~Si, X.~Qin, and P.~Zhang, ``A paradigm shift toward semantic communications,'' \emph{IEEE Communications Magazine}, vol.~60, no.~11, pp. 113--119, 2022.

\bibitem{rosso2020beyond}
P.~Rosso, D.~Yang, and P.~Cudr{\'e}-Mauroux, ``Beyond triplets: hyper-relational knowledge graph embedding for link prediction,'' in \emph{Proceedings of the web conference 2020}, 2020, pp. 1885--1896.

\bibitem{10233741}
Z.~Zhao, Z.~Yang, Y.~Hu, L.~Lin, and Z.~Zhang, ``Semantic information extraction for text data with probability graph,'' in \emph{2023 IEEE/CIC Int. Conf. Commun. China (ICCC Workshops)}, Aug. 2023.

\bibitem{wang2022performance}
Y.~Wang, M.~Chen, T.~Luo, W.~Saad, D.~Niyato, H.~V. Poor, and S.~Cui, ``Performance optimization for semantic communications: An attention-based reinforcement learning approach,'' \emph{IEEE Journal on Selected Areas in Communications}, vol.~40, no.~9, pp. 2598--2613, 2022.

\bibitem{zhou2022cognitive}
F.~Zhou, Y.~Li, X.~Zhang, Q.~Wu, X.~Lei, and R.~Q. Hu, ``Cognitive semantic communication systems driven by knowledge graph,'' in \emph{ICC 2022-IEEE International Conference on Communications}.\hskip 1em plus 0.5em minus 0.4em\relax IEEE, 2022, pp. 4860--4865.

\bibitem{zhang2016energy}
Y.~Zhang, H.-M. Wang, T.-X. Zheng, and Q.~Yang, ``Energy-efficient transmission design in non-orthogonal multiple access,'' \emph{IEEE Transactions on Vehicular Technology}, vol.~66, no.~3, pp. 2852--2857, 2016.

\bibitem{rimoldi1996rate}
B.~Rimoldi and R.~Urbanke, ``A rate-splitting approach to the gaussian multiple-access channel,'' \emph{IEEE Transactions on Information Theory}, vol.~42, no.~2, pp. 364--375, 1996.

\bibitem{clerckx2016rate}
B.~Clerckx, H.~Joudeh, C.~Hao, M.~Dai, and B.~Rassouli, ``Rate splitting for mimo wireless networks: A promising phy-layer strategy for lte evolution,'' \emph{IEEE Communications Magazine}, vol.~54, no.~5, pp. 98--105, 2016.

\bibitem{liu2020rate}
H.~Liu, T.~A. Tsiftsis, K.~J. Kim, K.~S. Kwak, and H.~V. Poor, ``Rate splitting for uplink noma with enhanced fairness and outage performance,'' \emph{IEEE Transactions on Wireless Communications}, vol.~19, no.~7, pp. 4657--4670, 2020.

\bibitem{clerckx2024multiple}
B.~Clerckx, Y.~Mao, Z.~Yang, M.~Chen, A.~Alkhateeb, L.~Liu, M.~Qiu, J.~Yuan, V.~W.~S. Wong, and J.~Montojo, ``Multiple access techniques for intelligent and multi-functional {6G}: Tutorial, survey, and outlook,'' 2024.

\bibitem{9145189}
Z.~Yang, J.~Shi, Z.~Li, M.~Chen, W.~Xu, and M.~Shikh-Bahaei, ``Energy efficient rate splitting multiple access (rsma) with reconfigurable intelligent surface,'' in \emph{2020 IEEE International Conference on Communications Workshops (ICC Workshops)}, 2020, pp. 1--6.

\bibitem{9382277}
Y.~Mao, E.~Piovano, and B.~Clerckx, ``Rate-splitting multiple access for overloaded cellular internet of things,'' \emph{IEEE Transactions on Communications}, vol.~69, no.~7, pp. 4504--4519, 2021.

\bibitem{10038476}
B.~Clerckx, Y.~Mao, E.~A. Jorswieck, J.~Yuan, D.~J. Love, E.~Erkip, and D.~Niyato, ``A primer on rate-splitting multiple access: Tutorial, myths, and frequently asked questions,'' \emph{IEEE Journal on Selected Areas in Communications}, vol.~41, no.~5, pp. 1265--1308, 2023.

\bibitem{9451194}
B.~Clerckx, Y.~Mao, R.~Schober, E.~A. Jorswieck, D.~J. Love, J.~Yuan, L.~Hanzo, G.~Y. Li, E.~G. Larsson, and G.~Caire, ``Is noma efficient in multi-antenna networks? a critical look at next generation multiple access techniques,'' \emph{IEEE Open Journal of the Communications Society}, vol.~2, pp. 1310--1343, 2021.

\bibitem{9831440}
Y.~Mao, O.~Dizdar, B.~Clerckx, R.~Schober, P.~Popovski, and H.~V. Poor, ``Rate-splitting multiple access: Fundamentals, survey, and future research trends,'' \emph{IEEE Communications Surveys \& Tutorials}, vol.~24, no.~4, pp. 2073--2126, 2022.

\bibitem{zhou2021rate}
G.~Zhou, Y.~Mao, and B.~Clerckx, ``Rate-splitting multiple access for multi-antenna downlink communication systems: Spectral and energy efficiency tradeoff,'' \emph{IEEE Transactions on Wireless Communications}, vol.~21, no.~7, pp. 4816--4828, 2021.

\bibitem{zhao2023semantic}
Z.~Zhao, Z.~Yang, Q.-V. Pham, Q.~Yang, and Z.~Zhang, ``Semantic communication with probability graph: A joint communication and computation design,'' in \emph{2023 IEEE 98th Vehicular Technology Conference (VTC2023-Fall)}.\hskip 1em plus 0.5em minus 0.4em\relax IEEE, 2023, pp. 1--5.

\bibitem{mao2018rate}
Y.~Mao, B.~Clerckx, and V.~O. Li, ``Rate-splitting multiple access for downlink communication systems: Bridging, generalizing, and outperforming sdma and noma,'' \emph{EURASIP journal on wireless communications and networking}, vol. 2018, pp. 1--54, 2018.

\end{thebibliography}

\end{document}